\documentclass[10pt]{article}
\usepackage{cell}
\usepackage{amsmath}
\usepackage{amssymb}

\usepackage{graphicx}

\usepackage{cite}
\usepackage{rotating}
\usepackage{color} 
\usepackage{multibib}
\newcites{Supp}{Supplementary References}

\usepackage[labelfont=bf,labelsep=period,justification=raggedright]{caption}

\makeatletter
\renewcommand{\@biblabel}[1]{\quad#1.}
\makeatother

\newcommand{\fig}[1]{Figure \ref{#1}}
\newcommand{\tab}[1]{Table \ref{#1}}

\newcommand{\usec}{[1/s]}

\newcommand{\umol}{[$\mu$M]}

\newcommand{\ti}{\kern -.02em\lower .6ex\hbox{\~{}}\kern .04em}
\begin{document}

% Title must be 100 characters or less
\begin{flushleft}
{\Large \bf A computational analysis of the dynamic roles of talin, Dok1, and PIPKI for integrin activation' }
\vskip2em
Florian Geier$^{1,2}$, 
Georgios Fengos$^{1}$, 
Dagmar Iber$^{1,\ast}$
\vskip2em
1 ETH Z\"urich, Department of Biosystems Science and Engineering (D-BSSE), Mattenstrasse 26, 4058 Basel, Switzerland; 2 new address: Biozentrum, Klingelbergstrasse 70, 4056 Basel, Switzerland
\vskip2em
$\ast$ E-mail: Corresponding dagmar.iber@bsse.ethz.ch
\vskip2em % up to 5 keywords
Keywords: Integrin Activation, Ensemble Modeling
\vskip0.5em % up to 50 characters
Running title: Modeling Integrin Activation 
\vskip0.5em
Subject Categories: Signal Transduction, Computational methods
\end{flushleft}

\newpage
% not more than 175 words
\section*{Abstract}
Integrin signaling regulates cell migration and plays a pivotal role in developmental processes and cancer metastasis. Integrin signaling has been studied extensively and much data is available on pathway components and interactions. Yet the data is fragmented and an integrated model is missing. We use a rule-based modeling approach to integrate available data and test biological hypotheses regarding the role of talin, Dok1 and PIPKI in integrin activation. The detailed biochemical characterization of integrin signaling provides us with measured values for most of the kinetics parameters. However, measurements are not fully accurate and the cellular concentrations of signaling proteins are largely unknown and expected to vary substantially across different cellular conditions.  By sampling model behaviors over the physiologically realistic parameter range we find that the model exhibits only two different qualitative behaviours and these depend mainly on the relative protein concentrations, which offers a powerful point of control to the cell. Our study highlights the necessity to characterize model behavior not for a single parameter optimum, but to identify parameter sets that characterize different signaling modes.

\newpage
\section*{Introduction}

Cell migration is a carefully regulated process that is essential for embryonic development and life \cite{Horwitz2003}. As the cell moves adhesion complexes form and dissolve. Key molecules in such focal adhesions are integrins, large membrane-spanning molecules that bind to ligands outside the cell and a variety of regulatory proteins inside the cell \cite{Arnaout2005,Banno2008,Campbell2008}. Integrins are allosteric proteins that can respond to extracellular and intracellular stimuli and change their affinity for ligand \cite{Hynes2002}. The two extreme conformations, an ÔopenÕ and a ÔclosedÕ one, bind ligand with maximal and minimal affinity respectively. The extracellular conformational changes are accompanied by movements of the intracellular domains which lead to a separation of the integrin tails \cite{Hynes2002}. Binding of ligand shifts the equilibrium to the active ÔopenÕ conformation. The separated integrin tails can then bind further signaling proteins and link to the cytoskeleton \cite{Harburger2009}. Intracellular activators such as talin and kindlins can also trigger integrin activation, a phenomenon that is referred to as Ôinside-out signalingÕ \cite{Moser2009}. Ligand-dependent Ôoutside-inÕ and signaling-dependent Ôinside-outÕ signalling are no separate processes; ligand binding leads to the activation of intracellular proteins that can, in principle, feed back on integrin activation. In fact recent experiments show that binding of talin to the cytoplasmic tails is essential for ligand-dependent integrin activation \cite{Zhang2008}. In the absence of talin, interaction with ligand leads only to a transient activation of downstream signaling and cells fail to adhere to the substrate \cite{Zhang2008}. 

Talin binds to the integrin beta-tail and stabilizes the active, open integrin conformation \cite{Tadokoro2003}. Most cellular talin is unavailable for integrin binding because of self-interactions between the PTB binding region and a tail region  \cite{Goksoy2008}. These inhibitory interactions can be relieved by binding of the lipid PIP2  \cite{Martel2001, Goksoy2008}. PIP2 is produced by type I phosphatidylinositol phosphate kinase-$\gamma$661 (PIPKI) and recruitment of PIPKI to focal contacts requires talin binding \cite{DiPaolo2002, Ling2003,  dePereda2005}. Ligand-bound integrins can stimulate the activity of PIPKI by enabling Src-mediated phosphorylation \cite{Ling2002,Ling2003}. Src kinase binds to beta-3 integrin tails \cite{Arias-Salgado2005,Arias-Salgado2003} and ligand-dependent clustering of integrins has been suggested to trigger Src auto-transphosphorylation. Integrin activation may thus trigger a positive feedback loop in that activation of Src kinases and PIPKI-dependent talin activation and recruitment enhances integrin activation. However, the architecture of this positive feedback loop is further complicated by the observed competition between integrin tails and PIPKI for talin binding \cite{Barsukov2003, dePereda2005, Goksoy2008}. Thus Src-dependent phosphorylation of PIPKI enhances the binding of talin and PIPKI, while Src-dependent phosphorylation of integrin beta-tails reduces their affinity for talin and increases their affinity for other competing signaling protein, i.e. Dok1 \cite{Ling2003, Oxley2008}. The latter effect has been coined Òintegrin phosphorylation switchÓ and has been suggested to induce a temporal switching from talin-dependent to Dok1-dependent integrin signaling. However, since only talin but not Dok1 stabilizes the open, active integrin conformation \cite{Wegener2007} it is unclear whether this switching is self-limiting and whether it can confer a switch in downstream signaling. The regulatory system is remarkably sensitive to the concentration of PIPKI: both a lower and a higher concentration impede talin recruitment and cell spreading \cite{Ling2002}. Does this help PIPKI to fulfill a dual role in first supporting integrin activation and then terminating integrin activation in a competition for talin \cite{Oxley2008}? Questions remain also regarding the exact mechanism as well as the purpose of these feedbacks. In particular, considering that ligand binding appears to be sufficient to trigger rapid and maximal integrin and Src activation \cite{Zhang2008} and that PIPKI has been suggested to sequester talin at a later stage \cite{Oxley2008} it is unclear why talin would be necessary for sustained integrin activation. 

Such questions are difficult to address by verbal reasoning alone. Mathematical modeling can help to integrate available isolated experimental information into a single model and permits the efficient analysis and comparison of model alternatives. Given the many states and complexes that need to be considered the dynamics of integrin activation can be best captured by a rule-based modeling approach \cite{Faeder2009}. Rule-based modeling permits the use of available information about complex protein-protein interactions in a precise and compact way. It is thus a convenient tool to construct a large and complex signaling network from a set of biochemical reaction rules \cite{Nag2010}. In decades of detailed biochemical analysis of proteins and their interactions most relevant rate and equilibrium constants have been measured. Yet, measurements are not fully accurate and the cellular concentrations of signaling proteins are largely unknown and expected to vary substantially across different cellular conditions.  To account for this level of uncertainty, we devise an ensemble modeling approach to characterize the biologically feasible dynamic range of integrin signaling. We realize this approach by a parameter sampling strategy. By integrating the available biochemical information and employing an ensemble modeling approach we address the following questions: (i) Can our model recapitulate both modes of signaling: outside-in and inside-out? What are the respective parameters/control points for both signaling modes? (ii) What is the role and mechanism of mutual talin/PIPKI membrane recruitment during signaling? (iii) What is the role of integrin phosphorylation and Dok1 recruitment during integrin activation? Is there an integrin phosphorylation switch as hypothesized previously \cite{Oxley2008}?

% Results and Discussion can be combined.
\section*{Results}

\subsection*{Model setup and parameterization} 

Integrin activity is regulated by the interaction with the extra-cellular matrix (outside-in signaling) and intra-cellular cytoskeletal adaptors (inside-out signaling). We use a rule-based modeling approach, in order to capture the complexity of integrin signaling. This approach allows us to model integrin signaling compactly by a set of 30 biochemical reactions rules, which are summarized graphically in \fig{fig:model} and are described in detail in the Supplementary Material. The model considers six components: integrins (INT), ligand (L), talin (TAL), PIP kinase (PIPKI), Src kinase (SRC) and Dok1 (DOK). The rules describe molecular interactions, state transitions (such as protein phosphorylation), and translocation between cytoplasm and membrane. We consider these two compartments because recruitment of proteins to the membrane enhances their local concentration. By considering two compartments we can capture this effect without having to include space explicitly in our simulation. We do not model the production and turnover of proteins, as these happen mostly on a time scale different from integrin activation. \fig{fig:model}A depicts the possible interactions (solid lines) and translocations between cytoplasmic and membrane compartments (dashed lines). Binding sites are specified by a circle, while interactions are depicted as solid lines connecting circles. Interactions which are competitive and cannot occur at the same time are shown as half-filled circles. Binding sites whose affinity is regulated by phosphorylation, like the NPxY binding motif on integrin tails, have an additional state indicated (U or P in this case). Other phosphorylation motifs are shown as square boxes containing either U or P while conformational states are also given as square boxes with either an open (O) or closed (C) state. The corresponding rule number which encodes the interaction or state transition is given next to each link. All state transitions which do not reflect binding interaction or translocations are additionally shown in \fig{fig:model} B. Note that each rule encodes only the relevant biochemical context for its reaction to happen. It could therefore apply to many species at the same time. We have encoded the 30 reaction rules within the BNGL modeling language \cite{Faeder2009} and compiled them into a system of ODEs based on the assumption of mass-action kinetics. This results in a reaction network of 108 species connected by 456 fluxes. 

The basic dynamics of our model are as follows. Binding of ligand and/or talin stabilizes the open/active conformation of integrin (rule R2, rules R3a and R4a). In case of outside-in signaling integrin signaling is initiated by binding of ligand (rule R2 in \fig{fig:model} A). Most cellular talin is unable to bind integrins because of inhibitory self-interactions. Talin can be activated by PIP2. Since PIP2 is highly unstable, its production, decay and diffusion are not modeled explicitly. We rather assume that talin must be bound to membrane-bound, active PIPKI for activation (rule R11a) and only a small fraction of membrane bound talin can be activated independently of PIPKI (rule R11b). PIPKI is activated by SRC-mediated phosphorylation (rule R8). Src kinases are activated by trans auto-phosphorylation which in case of outside-in signaling  is most likely mediated by juxtaposed integrin-ligand complexes. We do not model juxtaposition of integrins explicitly but use ligand-bound integrins as a proxy for juxta-position of integrins (rule R7a). The second mode of integrin activation, inside-out signaling, 
depends on integrin-independent talin activation and recruitment to the membrane. The GTPase Rap1 appears to play an important role in the recruitment of talin \cite{Banno2008}. Moreover, cross-talk from other pathways may lead to Src kinase and subsequently PIPKI activation. Since direct talin recruitment by Rap1/RIAM provides a straight-forward mechanism for integrin activation we focus on the more intricate path via Src kinase activation by cross-talk from other pathways (rule R7b in \fig{fig:model} B). TAL, DOK and PIPKI can translocate between cytoplasm and membrane compartments (rule R16, R17a/b/c and R18) which further modulates integrin activity.

Given the large body of experimental literature on integrin signaling measured values could be obtained for almost all rate constants (\tab{tab:par}). Nevertheless, some uncertainty remains, partly because of inaccurate or missing measurements, 
but also because of natural variations, in particular in protein concentrations. Given the likely regulatory impact of different protein concentrations we sought to analyze the entire physiological plausible range. Thus, based on the experimental data we defined for each of the 33 model parameters the most likely value as well as the range in which the parameter value was likely to lie. Here we considerd three classes of uncertainty: parameters for which there is detailed biochemical data were explored over a 2-fold range, parameter values that were based on indirect or less reliable data were modulated over a 5-fold range and parameter values that were based on measured values for similar proteins were allowed to change by up to 10 fold. The maximal range to be explored was thus a 100-fold difference in either concentration or time-scales for one parameter. Larger changes are unlikely to be of relevance and may lead to effective model reductions due to scale separations. Model parameters, measured values and references and as well as the considered ranges which reflect parameter uncertainty are given in \tab{tab:par}.

\subsection*{Analysis of parameter uncertainty} 

We use a sampling strategy to evaluate the influence of parameter uncertainty. As initial conditions we always used a common ground state of the network, i.e. the state that is attained without external activators such as ligand or external Src kinase activation. We then sampled each parameter independently from an exponential distribution within the range specified in \tab{tab:par}. This approach achieves an even sampling of the orders of magnitude as encoded by the fold change. In principle, our sampling space is very large. There are $(2\times2)^{8}(2\times5)^{11}(2\times10)^{14} \sim 10^{34}$ possible fold-change combinations, making exhaustive sampling infeasible. However, we expect the space of qualitatively different model dynamics to be much smaller. In order to estimate the number of samples needed to obtain a comprehensive picture of the possible dynamic range of the model for the given parameter uncertainty we monitored the convergence of the mean of each observable and time point in dependence of the sample size. We obtain convergence for $10^5$ parameter samples ($CV<0.1$, see supplemental data). We note that this approach, in general, will not capture rare dynamical events. 

When we simulate ligand-dependent signaling with the $10^5$ different parameter combinations we notice that the level of biologically relevant output variables (i.e. open integrins, membrane-bound TAL, and TAL-bound integrins) varies strongly between simulations (\fig{fig:ensemble}). In spite of this large uncertainty we can clearly discern two time scales in the model behaviour: fast ligand-driven activation of integrins (occurring on the second-to-minute scale) and slow TAL membrane recruitment and TAL:INT and DOK:INT complex formation (occuring on the minute-to-hour time scale). However, apart from the time scales little can at first be said about the extent of integrin activation and its interaction with talin or Dok1 because 95 \% of the results cover more than 50\% of the possible dynamic range (\fig{fig:ensemble} B,D dotted lines). Similarly, DOK binding to integrin is highly variable when comparing its dynamic range (95 \% of the results cover 20\% of the dynamic range) to the low mean level of DOK:INT complex formation (less than 1\%)  (\fig{fig:ensemble} C). 

We wondered whether the impact of parameters on the biologically relevant observables would be correlated. We therefore analysed pairwise scatter plots for eight observables that best capture the biologically interesting behaviour of the model, using the  $10^5$ parameter samples underlying \fig{fig:ensemble}. The first two observables quantify the extend of DOK and TAL binding to integrins in response to ligand stimulation (outside-in signaling) (\fig{fig:criteria} A). The mutual dependence of TAL and PIPKI for membrane recruitment is captured by observables three and four (\fig{fig:criteria} B), while the level of integrin and PIPKI phosphorylation is quantified by observables 5 and 6 (\fig{fig:criteria} C). The last two observables characterize the extend of integrin activation in response to outside-in and inside-out signaling (\fig{fig:criteria} D). A formal definition of all eight criteria is given in the Materials and Methods section. As TAL and DOK bind to integrin tails in a competitive manner, we expect that TAL:INT and DOK:INT are exclusive states for most of the parameters sets. This is indeed confirmed by \fig{fig:criteria} A, which shows beside the large marginal dynamic ranges of TAL:INT and DOK:INT, that both complexes are negatively correlated throughout the parameter samples. Positive correlation is observed between the extent to which parameter sets enable outside-in and inside-out signaling (\fig{fig:criteria} D). Thus, whenever a parameter set achieves a high level of INT activation by outside-in signaling, it also allows for high levels of INT activation by SRC crosstalk in the absence of any ligand. 

To better understand how the eight criteria interlink, we picked the 2\% extreme cases of the distribution in \fig{fig:criteria} A, the parameter sets allowing either for high TAL:INT formation (red points) or high DOK:INT formation (blue points) and marked the corresponding parameter sets also in the other three scatterplots. In this way we find that TAL:INT complex formation and the ability for high levels of outside-in and inside-out signaling is strongly linked (compare red points \fig{fig:criteria} A and D). Also the level of INT phosphorylation is lower for parameters which achieve high levels of INT activation (i.e. open integrins) (\fig{fig:criteria} C), while the level of PIPKI activation is high (\fig{fig:criteria} B). Interestingly, parameters which favor DOK:INT complex formation do not allow for inside-out signaling (compare blue points \fig{fig:criteria} A and D). INT phosphorylation in this case is also limited but significantly higher than for the TAL:INT favoring parameter sets (\fig{fig:criteria} D). The distribution of PIPKI activity is rather broad (\fig{fig:criteria} B).  Also the membrane recruitment of PIPKI is in most cases independent of TAL as seen by the distribution of blue points in the lower part of the vertical axis in \fig{fig:criteria} B. Parameters which allow for high levels of INT activation have a clear tendency for TAL-dependent PIPKI recruitment in this case. Thus, it seems that for all eight criteria we can isolate two sets of strongly correlated parameters which can be linked to two distinct dynamical regimes of the model:

\begin{description}
\item[Group 1] is characterized by a strong potential to support INT activation either by outside-in or inside-out signaling, high levels of TAL:INT complex formation, a strong dependence of PIPKI recruitment on TAL, low levels of INT phosphorylation and high levels of PIPKI activation. 
\item[Group 2] is characterized by the inability to support inside-out signaling, high levels of DOK:INT complex formation, no dependency of PIPKI recruitment on the presence of TAL and higher levels of INT phosphorylation compared with group 1. 
\end{description}

We next sought to define the parameter ranges that would correspond to the two distinct model behaviours and identify those parameters that would affect the two model behaviours the most. To this end we selected the parameter samples that corresponded to each group and compared the distribution of the single parameters (\fig{fig:criteria} E, group 1 (red boxes) or group 2 (blue boxes)) with their original sampling distribution (\fig{fig:criteria} E, grey boxes). We reasoned that sensitive parameters would be sampled from a strongly restricted range. The extent of the deviation was computed as the maximal difference between the cumulative distribution function (CDF) of a parameter and the CDF of its uniform sampling distribution. The absolute maximal difference, also called the Kolmogorov-Smirnov test statistics, was used to rank parameters for their relative influence on the group behavior (\fig{fig:criteria} F, red bars for group 1, blue bars for group 2, both versus the uniform sampling distribution, and gray bars for the difference between group 1 and group 2). The strongest deviation from the uniform sampling distribution is seen in the parameters for the relative total protein concentrations. Group 1, which is characterized by the capability of inside-out signaling, has a clear preference for higher TAL levels compared with the levels of integrin, DOK and PIPKI. Additionally, the PIPKI-dependent TAL activation rate is enhanced, while the PIPKI-independent activation is diminished and the binding and unbinding of TAL and PIPKI is shifted to faster time scales. Together this indicates a crucial role for TAL and PIPKI for the behavior of group 1. The same parameters tend to be distributed differently in group 2. Here the distribution of the total concentration ratios is just the opposite with a lower TAL concentration in comparison to INT, DOK and PIPKI levels. Some of the kinetic parameters are of characteristic importance, too. Thus the time scales for the binding and unbinding of DOK to phosphorylated INT are much faster on average, further disfavouring the formation of TAL:INT complexes and allowing for higher levels of DOK to INT binding. Interestingly the time scales for ligand-integrin interaction are lower on average, suggesting that a more stable ligand INT interaction correlates with higher extents of DOK binding. This result can be understood in light of the particular biochemical constraints: ligands need to stabilize open INT for DOK to bind to integrin tails because DOK cannot stabilize INT in its open conformation \cite{Wegener2007}. Overall this analysis highlights that the qualitative behaviour of biological networks depends only on few parameters. Interestingly, the two different dynamic regimes uncovered by our analysis (group 1 and 2 parameter sets) are mainly defined by the relative total concentrations of TAL, DOK and PIPKI. Cells have powerful mechanisms in place to adjust relative protein concentrations. These thus provide excellent control points for cells to define its cellular dynamics. We thus hypothesize that integrin signaling operates in both regimes by altering the relative levels of TAL, DOK and PIPKI. In the following we investigate their biological role in more detail.
 
\subsection*{Integrin phosphorylation switch}

Detailed biochemical measurements have led to the proposition of an integrin phosphorylation switch \cite{Oxley2008}. Thus measurements revealed that Src-dependent phosphorylation of integrin beta-tails lowers the integrin-talin affinity by about 2-10-fold \cite{Ling2003,Oxley2008} while enhancing the affinity of Dok-1 for the same binding site some 400-fold \cite{Oxley2008}. Some experiments also reported an increased affinity of PIPKI for talin once PIPKI had been phosphorylated by Src kinases \cite{Ling2003, dePereda2005}. Src-dependent phosphorylation of integrin tails and PIPKI upon integrin activation was thus suggested to result in a time-dependent exchange of TAL for DOK on integrin tails and the sequestering of TAL in complexes with PIPKI, coined integrin phosphorylation switch \cite{Oxley2008}. We use our simulation to explore this hypothesis for the two groups of parameter sets identified above. Thus \fig{fig:intphos} A-D shows the extent of integrin phosphorylation and binding to DOK or TAL over time and dependent on the DOK/TAL ratio for one set of parameters characteristic for each group of parameter sets. The characteristic parameter set represents the mean value of all parameter samples belonging to each group \tab{tab:parS}. As expected, the level of TAL:INT exceeds the level of DOK:INT in group 1 (\fig{fig:criteria} A,B), and vice versa for group 2 ( \fig{fig:criteria} C,D). The level of integrin phosphorylation is higher for group 2 because in case of group 1 parameters TAL binding shields the  NPxY phosphorylation motif in integrin tails and thus prevents phosphorylation, while in case of group 2 parameters the unphosporylated NPxY motif is accessible and bound DOK protects the phosphorylated NPxY motif from dephosphorylation. In spite of strong integrin phosporylation for group 2 parameters we, however, do not observe a temporal switching from TAL:INT to DOK:INT association (\fig{fig:criteria} C). Instead, both complexes coexist for comparable levels of TAL and DOK in case of group 2 and balance in favor of TAL:INT in case of group 1 (\fig{fig:criteria} A,C).

Although there is no temporal phosphorylation switch, its is possible to switch between TAL and DOK bound forms of integrin by regulating the ratio of total concentrations of DOK and TAL (\fig{fig:intphos} B and D). For group 2 the switching point is reached at equal DOK and TAL total concentrations which is close to the likely physiological protein concentrations (dotted vertical  lines). For group 1 parameters on the other hand the switching point is reached at 1000-fold higher levels of DOK over TAL which is physiologically unrealistic. Thus, in group 1, TAL:INT complex formation seems robust to the DOK-to-TAL ratio, while for the parameter sets in group 2 physiological changes in the protein concentrations could lead to a switch. \fig{fig:intphos} B and D confirm our previous observation that the level of phosphorylated integrin negatively correlates with the total level of TAL and DOK-bound integrins.

While we can exclude a temporal switch in DOK and TAL binding of integrin tails we note a temporal switch in integrin phosphorylation for group 1 parameters (\fig{fig:intphos} A). Thus as talin accumulates at the membrane and binds to integrin tails the extent of integrin phosphorylation decreases after having reached an early maximum of about $60\%$ of phosphorylation. Integrin signaling is thus transient. To see whether this is a general behavior for group 1 parameters and wether its also hold for group 2 parameters we analysed the entire parameter sets belonging to either group ( \fig{fig:intphos} E). We used two measures for transient responses: sensitivity towards stimulation and the precision with which the extent of integrin activation (E), INT phosphorylation (F) and TAL:INT complex formation (G) return to pre-stimulation levels (see materials and methods section for details). Transient responses are characterized by high sensitivity and precision while sustained responses are characterized by high sensitivity and low precision. A low sensitivity would characterize unresponsive systems. We observe a clear separation for the two parameter sets in the scatter plots in \fig{fig:intphos} E-G. Thus group 1 parameter sets result in self-limited integrin phosphorylation; phosphorylation levels, however, do not return to base line but assume some intermediate level (i.e. $50\%$ phosphorylation, \fig{fig:intphos} A+F). Group 2 parameters on the other hand enable a very sensitive, sustained phosphorylation response. Even though integrin phosphorylation is necessary for downstream signaling, integrin activation is typically monitored as the extent to which integrins assume the open conformation. Interestingly, this alternative measure of integrin activation gives very different results (\fig{fig:intphos} E). Thus group 1 parameters lead to strong and sustained integrin activation (even though phosphorylation is self-limiting) while Group 2 parameters lead to intermediate (if sustained) integrin activation. In fact all simulated $10^5$ parameter combinations result in sustained levels of open, active integrins as indicated by the distribution of all parameter sets along the diagonal (\fig{fig:intphos} E). As expected from the previous analysis, the establishment and temporal dynamics of TAL:INT signaling correlates well with integrin activity, showing high sensitivity and sustained response for group 1 parameters (\fig{fig:intphos} G). Phosphorylation and downstream signaling can thus be independently controlled from the conformational change that initiates integrin activation.

\subsection*{Mutual regulation of TAL and PIPKI membrane recruitment}

TAL and PIPKI are cytoplasmic proteins that need to be enriched at the membrane to serve their function in integrin activation. However, it has been difficult to explain how they become enriched at the membrane as both bind to membrane lipids at most weakly. In the following we focus our analysis on group 1 parameters since only these permit strong INT:TAL complex formation (\fig{fig:criteria} A). \fig{fig:criteria} B suggested that PIPKI and talin mutually enhance each others membrane recruitment, and indeed for very low levels of either TAL or PIPKI there is no enhanced membrane recruitment (\fig{fig:memrec} A and B). TAL enhances PIPKI recruitment by acting as a membrane adaptor that binds both the membrane and PIPKI. PIPKI, on the other hand, enhances TAL recruitment in the model because as a dimer \cite{Rao1998} it multimerizes TAL and thus enhances its affinity for the membrane. Moreover, it activates TAL by producing PIP2 and thus enables its binding to integrin tails. Since the PIPKI:TAL interaction does not depend on integrin activation both accumulate at the membrane to a considerable extent also in the absence of ligand (\fig{fig:memrec} C). Integrin activation then enhances recruitment, mainly by permitting INT:TAL complex formation. As observed previously these complexes only form minutes to hours after ligand addition even though open integrins emerge within seconds \cite{Ling2002}. 

The extent of integrin activation strongly depends on the total talin and PIPKI concentrations  (\fig{fig:memrec} A and B). Thus the fraction of active integrins increases as the talin concentration is increased and reaches near maximal levels at the likely physiological concentration of 10 $\mu$M (\fig{fig:memrec} A). Both the extent of integrin activation and talin recruitment to the membrane exhibit a bell-shaped dependency on the PIPKI concentration (\fig{fig:memrec} B). At low PIPKI concentrations too little scaffold is available for talin to be multimerized at the membrane while at very high concentrations all talin becomes sequestered in complexes with PIPKI and is no longer available for integrin activation. Similar as discussed for TAL/DOK there is thus no temporal but a concentration-dependent switching point for integrin activation. The bell-shaped model prediction is in agreement with the experimental observations: mild over-expression of PIPKI enhances talin recruitment while strong over-expression of PIPKI inhibits talin recruitment \cite{Ling2002} and leads to the dissolution of focal adhesions \cite{DiPaolo2002}. This experimental observation raised the question as to why physiological levels are suboptimal for talin recruitment. Based on our simulations we note that the curves for talin recruitment and integrin activation do not coincide. Maximal integrin activation is achieved for lower levels of PIPKI compared with the levels for maximal TAL recruitment. The mean value of group 1 parameters (which were selected for highest TAL:INT formation) coincides with maximal integrin activation (1 $\mu$M PIPKI, \fig{fig:memrec} B, dotted line). We thus propose that the signaling network was optimized to maximize integrin activation in response to ligand binding, and that this maximum is achieved when the extent of talin recruitment is sub-optimal as observed in experiments.

\subsection*{Outside-in and inside-out signaling}

Integrins are unusual receptors in that they can be activated both by external signal (outside-in) and internal signals (inside-out signaling). As discussed above both modes of signaling are observed only with group 1 parameters (\fig{fig:criteria} D) and we will therefore now focus on these parameter sets. Both signaling modes result in a similar extent of integrin activation yet the kinetics are very different. Thus outside-in signaling follows a biphasic time course with an early talin-independent phase and a later talin-dependent phase (\fig{fig:oiio} A, compare black solid and dashed lines) while inside-out signaling exhibits only the later talin-dependent signaling phase (\fig{fig:oiio} A, compare magenta solid and dashed lines). The increased talin dependency of inside-out signaling becomes apparent also in \fig{fig:oiio} B where we record steady state integrin activation in dependence of total talin levels for both signaling modes.   

\fig{fig:oiio} C illustrates the extent to which talin supports outside-in signaling at low ligand densities. High levels of open integrins appear already for low ligand levels but are absent in the TAL knock-out. As expected, SRC levels play a similar role for inside-out signaling as ligand levels for outside-in signaling (\fig{fig:oiio} D) since total SRC levels are a proxy for the amount of SRC which becomes activated by cross-talk signaling. Both INT activation curves are qualitatively similar, except that full INT activation is only achieved in the presence of high ligand concentrations. Interestingly, both curves show an intermediate range of high integrin activation ($L/K_D \sim 10^{-5} \mu$M, \fig{fig:oiio} C) and for inside-out signaling lower levels of SRC achieve highest INT activation (SRC $\sim 10^{-2} \mu$M, \fig{fig:oiio} D). In both cases, membrane levels of PIPKI increase with the amount of INT activation. Since SRC binds and sequesters PIPKI at the membrane, it indirectly attenuates the role of PIPKI in recruiting TAL. Thus, as observed in \fig{fig:memrec} B, there is an optimal level of PIPKI at the membrane for INT activation, which is lower and depends on the level of ligand and SRC. Despite this intermediate maximum, the model predicts high levels of integrin activation over a wide range of ligand and Src kinase concentrations. Thus, while PIPKI levels appear to be optimized for integrin activation, the signaling mechanism appears to be much more robust to variations in ligand and SRC levels, so that integrin signaling happens almost in a none-or-all fashion (\fig{fig:oiio} C, D).

\section*{Discussion}

We have developed a rule-based model to explore the roles of talin, PIPKI, and Dok1 in integrin activation and signaling. In spite of much detailed biochemical data a significant level of uncertainty remains because measurements are not fully accurate and the concentrations of signaling proteins are largely unknown and expected to vary substantially across different cellular conditions. It is generally infeasible to sample the entire physiological parameter space. We therefore addressed this challenge in modelling signaling pathways by restricting the sampled parameter space according to the quality of the available measurements. The width of the ranges was based on general rules as to the accuracy of different experimental methods. This restriction enabled us to systematically analyse model dynamics with only $10^5$ samples of parameter sets. 

Interestingly, in spite of the wide sampling we observed only two different qualitative behaviours: one group of parameter sets supported integrin activation both by outside-in or inside-out signaling, and was characterized by high levels of talin-integrin complex formation, a strong dependence of PIPKI recruitment on talin, low levels of integrin phosphorylation and high levels of PIPKI activation. The other group of parameter sets did not support inside-out signaling, and was characterized by high levels of Dok1-integrin complex formation, no dependency of PIPKI recruitment on the presence of talin and higher levels of integrin phosphorylation.

Importantly, the most decisive parameters were the relative cellular protein concentrations. Accordingly, the total concentrations of network components can serve as powerful control points to achieve distinct network dynamics. Thus for low talin concentrations we predict impaired inside-out signaling, yet much Dok1-integrin complexes form in response to outside-in signaling which will direct downstream signaling via recruitment of a range of proteins, including Ras-GAP, the adaptor protein Nck, the non-receptor tyrosine kinase Csk, and the phosphatase SHP2 \cite{Ling2005}. In the presence of high talin concentrations cells can trigger both outside-in and inside-out integrin signaling and talin can link integrins with the cytoskeleton. Our modeling suggests that under physiological outside-in signaling conditions both signaling platforms co-exist and that the equilibrium is mainly determined by the talin/Dok and talin/PIPKI ratios. 

Previous experimental observations led to the suggestion of an integrin phosphorylation switch. Biochemical measurements revealed that phosphorylation of integrins and PIPKI would lower the talin-integrin affinity and enhance PIPKI-talin binding \cite{Barsukov2003, dePereda2005, Goksoy2008}. Moreover, the Dok1-integrin affinity was enhanced some 400-fold  \cite{Ling2003, Oxley2008}. Based on our parameter screen we can dismiss the idea of a temporal phosphorylation switch. However, we note that a switch can be achieved if relative protein concentrations are changed. Such change in protein concentration must be the result of protein expression rather than membrane recruitment as the signaling induced recruitment dynamics were included in the model and are insufficient to trigger the switch. Membrane recruitment of talin and PIPKI itself poses an interesting conundrum in that both appear to depend on each other, yet both bind membrane lipids at most weakly. Based on our simulations we propose that PIPKI enhances talin recruitment to the membrane by offering a dimeric scaffold \cite{Rao1998} that multimerizes talin and thus enhances the effective talin-membrane affinity. Talin, on the other hand, recruits PIPKI to the membrane by binding to the scaffold. The concentration of PIPKI needs to be finely balanced for successful integrin activation. Thus, at low PIPKI concentrations there is insufficient scaffold for membrane recruitment and insufficient PIP2 production for efficient talin activation while at very high PIPKI concentrations active talin will be sequestered by the kinase and will not be available for binding to integrin tails. Therefore, the PIPKI level is one of the main control points for talin and integrin activation. Slight overexpression of PIPKI enhances talin recruitment in experiments  \cite{Ling2002}. While this suggested that the interactions are not optimized  for integrin activation, the simulation now reveals that maximal integrin activation is achieved already at a lower talin concentration, and the protein network thus appears to be optimized for integrin activation rather than talin recruitment. 

The simulation further revealed that the opening of the integrin conformation not necessarily leads to integrin phosphorylation.  Because binding shields the integrin phosphorylation motif, strong recruitment of either Dok1 or talin to integrin tails will counteract integrin phosphorylation. Therefore integrin phosphorylation is indirectly regulated by recruitment of talin and Dok1. We observe that conditions which enable high levels of Dok1 signaling also allow for a sensitive, sustained phosphorylation response while talin-dependent signaling allows only for transient levels of high phosphorylation. Since talin is the more stable signaling platform, its stable interaction with integrin tails does not allow for high levels of integrin phosphorylation but a sustained level of open integrins. Our modeling therefore suggest, that integrin activation and integrin phosphorylation serve as independent nodes in integrin signal propagation. Finally our simulation suggests that outside-in and inside-out signaling differ in the kinetics of the activation process. Thus while outside-in signaling is biphasic with a rapid talin-independent and a slow talin-dependent activation phase, there is only the late talin-dependent activation phase in inside-out signaling. 

It has long been recognized that model behaviours should be analysed not for a single measured parameter set alone, but  over a wider parameter range. Ensemble modeling strategies combined with various clustering algorithms have been used to reveal parameter sets that yield similar qualitative results \cite{Melke2006, Kuepfer2007}. The network studied here is unusual in that almost all kinetic parameters have been measured (Table \ref{tab:par}). However, in agreement with previous studies we notice that the qualitative model behaviour is not particular sensitive to most kinetic parameters, somethings that has been referred to as sloppiness \cite{Gutenkunst2007}. More important than the kinetic parameters are the protein concentrations which can differ between cells and conditions and which may give rise to heterogeneous functionality on the single cell level \cite{Spencer2009}. The large impact of protein concentrations provides cells with a powerful level of control and enables cells to assume to different modes of signaling in the context of our model here.

Overall, our study illustrates how detailed biochemical measurements together with a thorough computational analysis can be used to reveal the qualitative behaviours of biological signaling networks. Without biological data the space of possible parameter combinations is too large to be explored. Yet without the computational model it is impossible to integrate biological knowledge to grasp the dynamic range of biological networks and to estimate the possible impact of parameter variations under different cellular conditions.

% You may title this section "Methods" or "Models". 
% "Models" is not a valid title for PLoS ONE authors. However, PLoS ONE
% authors may use "Analysis" 
\section*{Materials and Methods}

\paragraph{Model setup} We use a rule-based approach to explore the roles of talin, Dok1, and PIPKI in integrin signaling dynamics \cite{Faeder2009}. Rule based modeling associates a pattern with a given biochemical reaction. The pattern specifies all relevant species for which the reaction is applicable. By that, rule based modeling gives a compact description of a biochemical reaction system and avoids redundant information, which may arise by the combinatorial complexity of most signaling systems. Our integrin signaling model consists of 6 molecule types and 29 reactions rules which were translated into a set of 108 species and 454 individual reactions using BioNetGenerator \cite{Faeder2009}. There are two reaction compartments: plasma membrane and cytoplasm. The membership to a certain reaction compartment was specified by molecule states and the exchange fluxes between compartments were adjusted on a per species basis in order to account for different compartment volumes (see Supplement for details). See Supplemental Data for a complete model description and the set of reaction rules. Time course and steady state solutions of the set of ODEs were computed numerically with the ode15s integrator of MATLAB. The ground state (GS) is defined as the steady state solution in the absence of any ligand or SRC cross-talk activation. It served as the initial condition for all simulations of integrin activation. Outside-in (OI) signaling was simulated by setting $L_{tot} = 15$, while inside-out (IO) signaling was achieved by setting $L_{tot} = 0$ and switching on rule 7b which leads to an integrin independent SRC activation.

\paragraph{Parameterization and treatment of uncertainty} We collected parameter values for all biochemical reactions from the literature and  parameterized the model according to \tab{tab:par} (see Supplement for details). Due to the heterogeneity of the biochemical data/essays and intrinsic uncertainty in reaction rate constants, we additionally defined a biologically plausible range for each rate constant. We sampled parameters within the ranges defined by the fold changes given in \tab{tab:par} in order to characterize the possible model behavior within the range of parameter uncertainty. Parameters were sampled uniformly on a $\log_{10}$ scale within the range $k_0/f \le k \le k_0f$, where $f$ is the fold change and $k_0$ the basal parameter value, both stated in \tab{tab:par}. In order to evaluate the number of parameter samples needed to cover the dynamical range of the model, we checked the convergence of the time-dependent mean of all model observables by a blocking procedure. Convergence was measured by the coefficient of variation and was achieved for approximately $10^5$ samples ($CV<0.1$). 

\paragraph{Definition of output criteria} We use the following eight criteria to characterize model behavior.
\begin{enumerate}
\item DOK:INT formation in response to stimulus:
\begin{eqnarray*}
C_{1} & = &  (DOK:INT_{OI} - DOK:INT_{GS})/INT_{tot} 
\end{eqnarray*}
\item TAL:INT formation in response to stimulus:
\begin{eqnarray*}
C_{2} & = &  (TAL:INT_{OI} - TAL:INT_{GS})/INT_{tot} 
\end{eqnarray*}
\item PIPKI-dependent TAL recruitment: 
\begin{eqnarray*}
C_{3} & = &  (TAL_{OI}^{Mem} - TAL_{Pipki}^{Mem})/TAL_{tot} 
\end{eqnarray*}
\item TAL-dependent PIPKI recruitment:
\begin{eqnarray*}
C_{4} & = &  (PIPKI_{OI}^{Mem} - PIPKI_{Tal}^{Mem})/PIPKI_{tot} 
\end{eqnarray*}
\item PIPKI phosphorylation in response to stimulus:
\begin{eqnarray*}
C_{5} & = &  (PIPKI_{OI}^{Phos} - PIPKI_{GS}^{Phos})/PIPKI_{tot} 
\end{eqnarray*}
\item INT phosphorylation in response to stimulus:
\begin{eqnarray*}
C_{6} & = &  (INT_{OI}^{Phos} - INT_{GS}^{Phos})/INT_{tot} 
\end{eqnarray*}
\item Integrin activation by outside-in signaling:
\begin{eqnarray*}
C_{7} & = &  (INT_{OI} - INT_{GS})/INT_{tot} 
\end{eqnarray*}
\item Integrin activation by inside-out signaling:
\begin{eqnarray*}
C_{8} & = &  (INT_{IO} - INT_{GS})/INT_{tot} 
\end{eqnarray*}
\end{enumerate}
Here, $Pipki$ and $Tal$ refers to the steady state of the PIPKI or TAL knock-out after ligand application. $Mem$ indicates the membrane bound forms and $Phos$ the phosphorylated forms. We additionally defined the sensitivity $S$ and the precision $P$ of an observable $Y$ as
\begin{eqnarray*}
% sensitivity
S = \left|(Y^{Max} - Y_{GS})/Y_{tot}\right|
\end{eqnarray*}
and
\begin{eqnarray*}
% precision
P = \left| (Y_{OI} - Y_{GS})/Y_{tot} \right|^{-1}.
\end{eqnarray*}
Here, $Max$ refers to the maximum of $Y(t)$ for $0 \le t \le 4$h.

\paragraph{Characterization of parameter classes} Based on the first two criteria, we selected the 2000 parameter sets (2\%) which achieved the highest levels of TAL:INT (group 1) and DOK:INT (group 2). In order to identify the relevant parameters which characterize both groups, we further compared the distribution of group 1 and group 2 parameters set with their corresponding uniform sampling distributions and among each other. We computed the empirical cumulative distribution function ($CDF(x)$) of parameter $p$ for the sampling distribution and in each of the two group distributions. We rank parameters for their influence on group behavior based on the maximal deviation between all three CDFs (i.e., the Kolmogorov-Smirnov test statistics $D$).  
\begin{eqnarray*}
    D_{g_1}^p = \sup_x \left|CDF^p_{g_1}(x)-CDF^p_s(x)\right| \\
    D_{g_2}^p = \sup_x \left|CDF^p_{g_2}(x)-CDF^p_s(x)\right| \\
    D_{g_{1/2}}^p = \sup_x \left|CDF^p_{g_1}(x)-CDF^p_{g_2}(x)\right| 
\end{eqnarray*}  
Here subscripts specify the group ($g_1,g_2$) and sampling distributions ($s$). 

% Do NOT remove this, even if you are not including acknowledgments
\section*{Acknowledgments}
The project was financially supported by grants 51RT-0-126008 (InfectX) and an iPhD grant from SystemsX.ch, the Swiss Initiative for Systems Biology.

%\section*{References}
% The bibtex filename
\bibliographystyle{/Users/iberd/MyDocuments/Academia/Publications/Bibliography/msb}
\bibliography{integrin.bib}

\pagebreak 

\section*{Figure Legends}

\begin{figure}[h!]
\begin{center}
\end{center}
\caption{{\bf A rule-based model of integrin signaling.} (A) Contact and localization map. (B) Additional state transition rules. The model considers six molecules: ligand (L), integrin (INT), Dok (DOK), talin (TAL), Src kinase (SRC) and pip kinase (PIPKI). Each molecule has a set of binding site and a set of possible states. Reactions take place within three compartments denoted as extracellular domain (EC), plasma membrane (PM) and cytoplasm (CP). DOK, TAL and PIPKI can translocate between the CP and PM compartments, while INT and SRC only reside within the PM compartment. L is extracellular. Possible binding reactions (annotated by the rule number) and involved binding site are indicated. Note, that some binding sites are competitive, e.g., the NPxY motif on INT can be either DOK or TAL bound. For a formal definition of the reaction rules see Supplemental Data.}
\label{fig:model}
\end{figure}

\begin{figure}[h!]
\begin{center}
\end{center}
\caption{{\bf Integrin signaling dynamics within biologically feasible parameter ranges.} Average time courses of integrin activation (A), TAL membrane recruitment (B), DOK bound INT (C) and TAL bound INT (D). Median (black line), 50 \% (dashed line) and 95 \% (dotted line) data intervals. To better visualize the temporal order of integrin activation, time is represented on a log-scale. Parameters are varied according to the ranges given in \tab{tab:par}. All simulations were performed with $L/K_D$ = 50. Results are given for $10^5$ samples.}
\label{fig:ensemble}
\end{figure}

\begin{figure}[h!]
\begin{center}
\end{center}
\caption{{\bf Two groups of qualitatively different model behavior.} (A-D) Scatter plots of eight criteria used to evaluate model behavior. (A) Ligand-induced DOK:INT versus TAL:INT complex formation. The 2\% extreme ends of the distribution are colored either red (TAL:INT) or blue (DOK:INT) and reflect two qualitatively different model behaviors. Both groups of samples are mapped into panel B-D to visualize the dependencies between criteria. (E) Box plots of the parameter samples of each group. Ranges of the uniform sampling distributions, as stated in \tab{tab:par}, are indicated by grey boxes. (F) Maximal difference in the cumulative distributions between sampling parameters and selected parameter sets: TAL:INT (blue), DOK:INT (red) and the difference between both groups (grey).}
\label{fig:criteria}
\end{figure}

\begin{figure}[h!]
\begin{center}
\end{center}
\caption{{\bf Integrin phosphorylation and complex formation for average group 1 and group 2 parameter values.} Both groups are colored according to \fig{fig:criteria}: red for group 1, (panel A and B) and blue for group 2 (panel A and C). (A,C) Time course of integrin phosphorylation as well as DOK:INT, TAL:INT and TAL:PIPKI complex formation. (B,D) Steady state levels of the corresponding species as a function of the ratio of DOK/TAL. Vertical line: mean DOK/TAL ratio in both groups. (E-G) Sensitivity and precision, as defined in Material and Methods section, of open INT (E), phosphorylated INT (F) and TAL:INT complex (G).}
\label{fig:intphos}
\end{figure}

\begin{figure}[h!]
\begin{center}
\end{center}
\caption{{\bf Membrane recruitment of TAL and PIPKI through ligand induced integrin activation.} (A) Fraction of membrane-recruited PIPKI and open INT as a function of total TAL levels. (B) Fraction of membrane-recruited TAL and open INT as a function of total PIPKI levels. (C) Time course of membrane recruitment of TAL, PIPKI and TAL/PIPKI complex. All results are based on the parameter set characterizing group 1 behavior. Vertical lines: group 1 TAL (A) and PIPKI (B) levels.}
\label{fig:memrec}
\end{figure}

\begin{figure}[h!]
\begin{center}
\end{center}
\caption{{\bf Integrin activation by outside-in and inside-out signaling.} (A) Time course of INT activation. Outside-in (OI) signaling (black), inside-out (IO) signaling (magenta), wild type (solid), TAL knock out (dashed). (B) Fraction of open INT in dependence of total TAL. (C) Fraction of open INT in dependence of $L/K_D$. (D) Integrin activation by inside-out signaling as a function of SRC levels. Vertical line: mean group 1 parameter values.}
\label{fig:oiio}
\end{figure}

\clearpage
\newpage

\pagebreak

\section*{Figures}

\begin{center}
\includegraphics[width=0.9\textwidth]{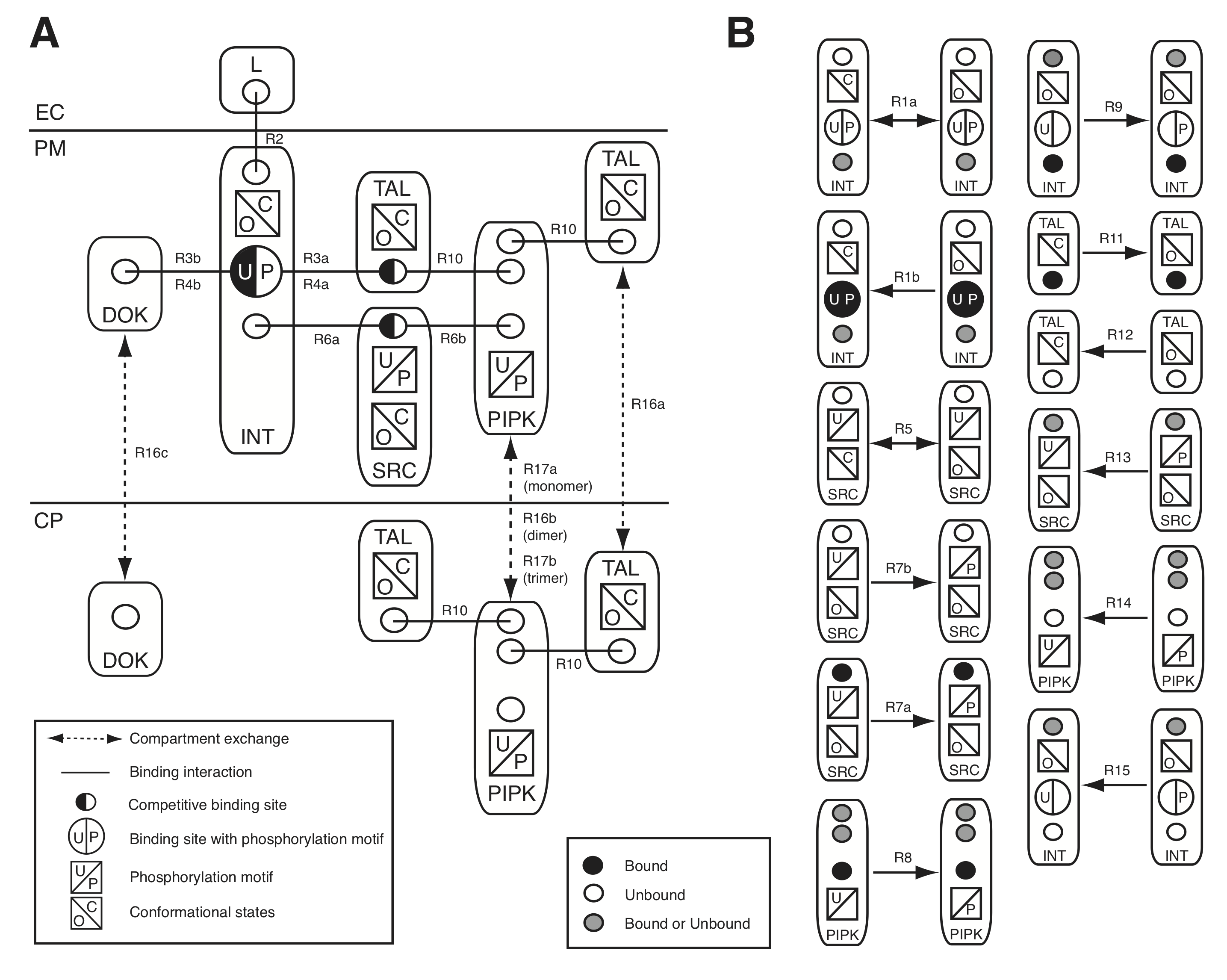}
\end{center}
{\bf Figure 1} 

\newpage 

\begin{center}
\includegraphics[width=0.9\textwidth]{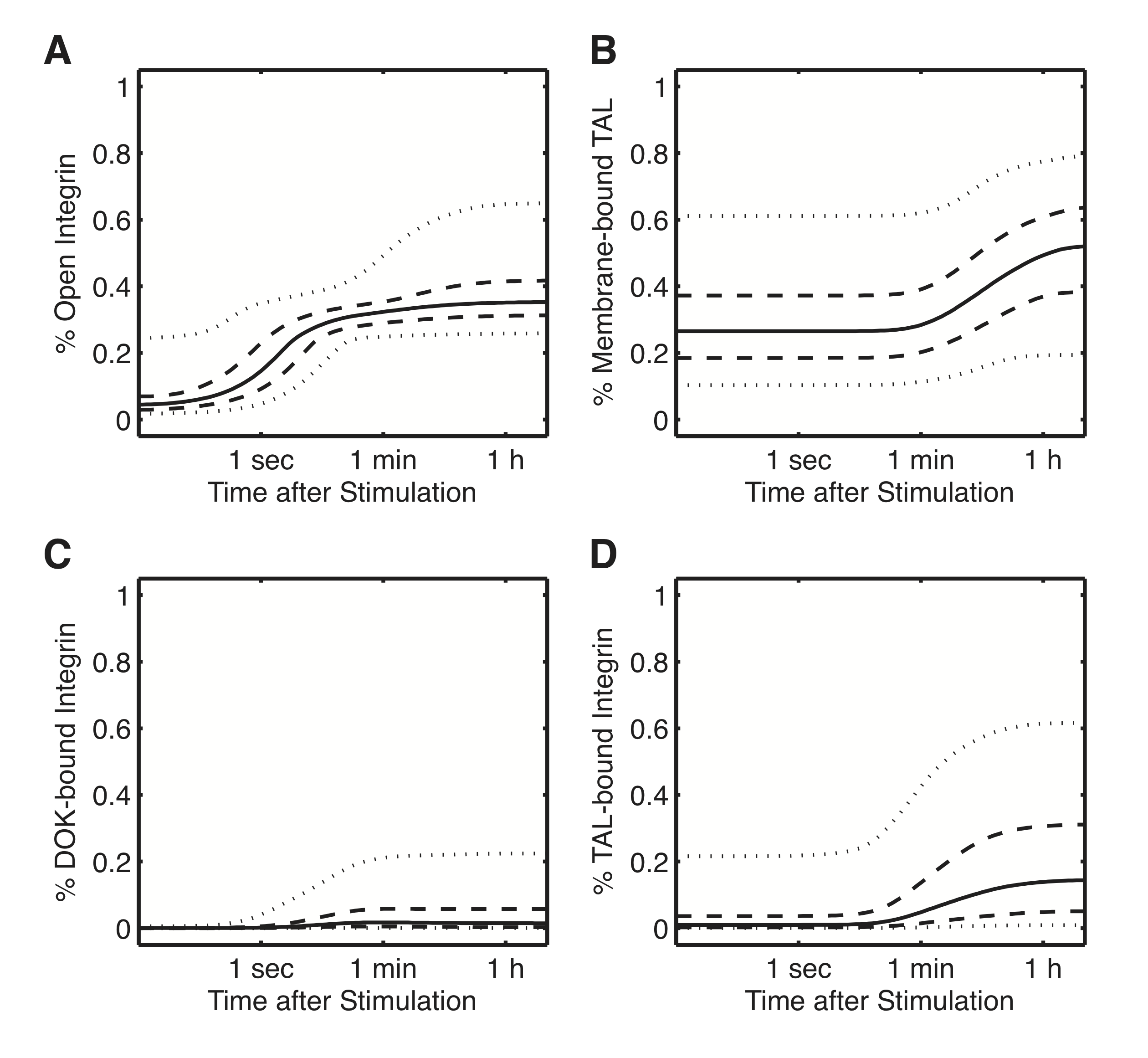}
\end{center}
{\bf Figure 2}

\newpage 

\begin{center}
\includegraphics[width=0.8\textwidth]{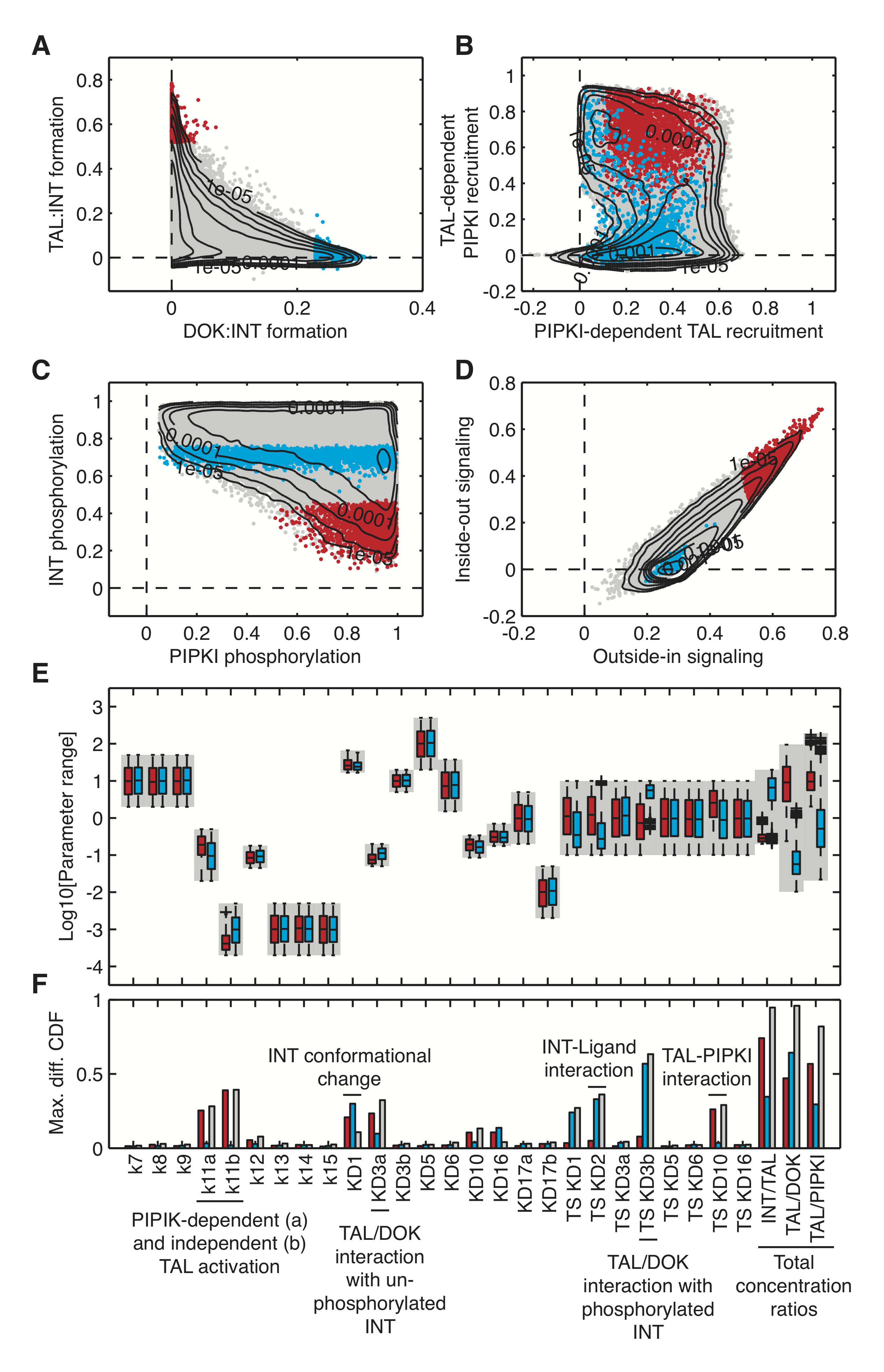}
\end{center}
{\bf Figure 3}

\newpage 

\begin{center}
\includegraphics[width=1\textwidth]{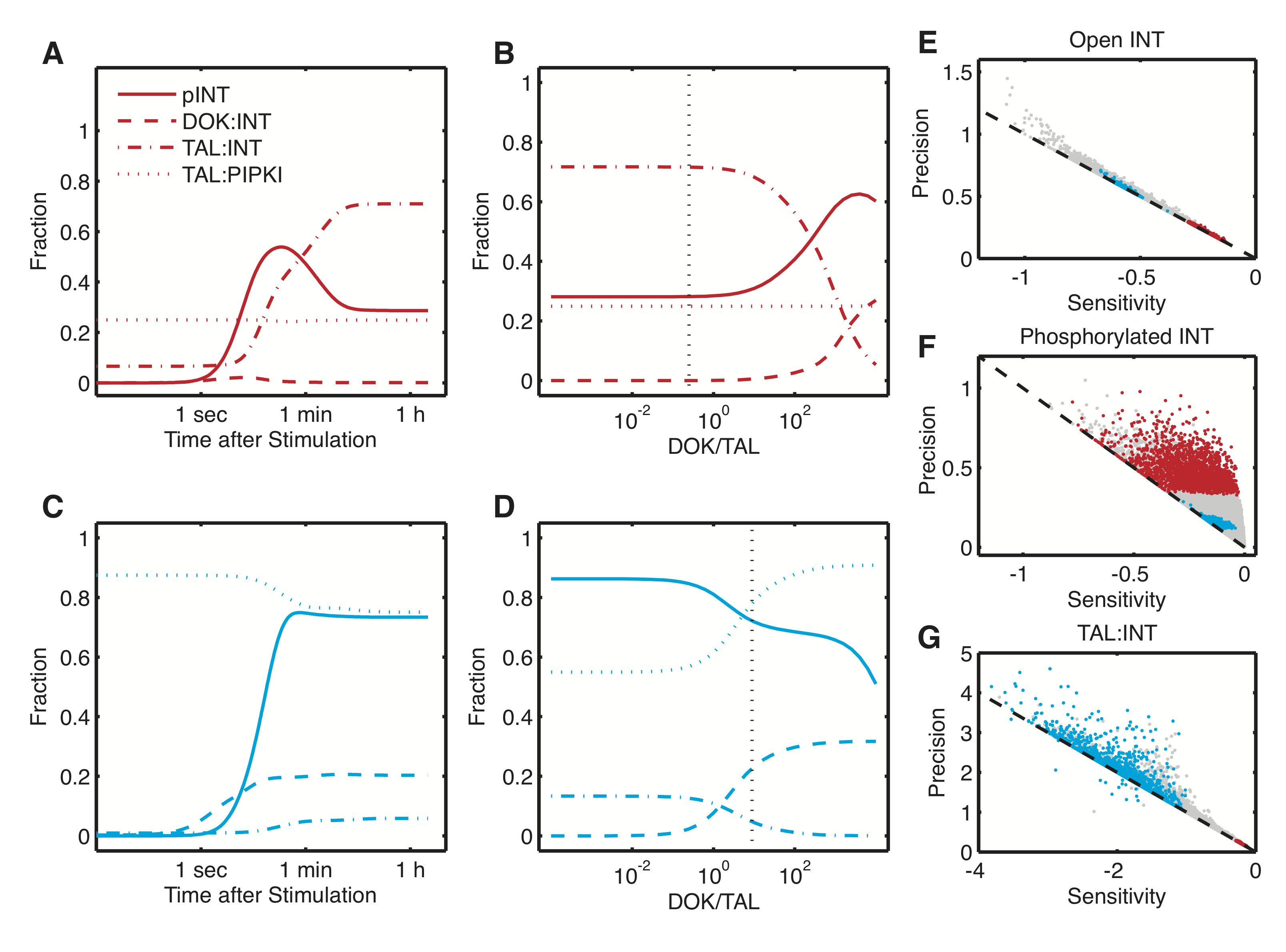}
\end{center}
{\bf Figure 4}

\newpage 

\begin{center}
\includegraphics[width=1\textwidth]{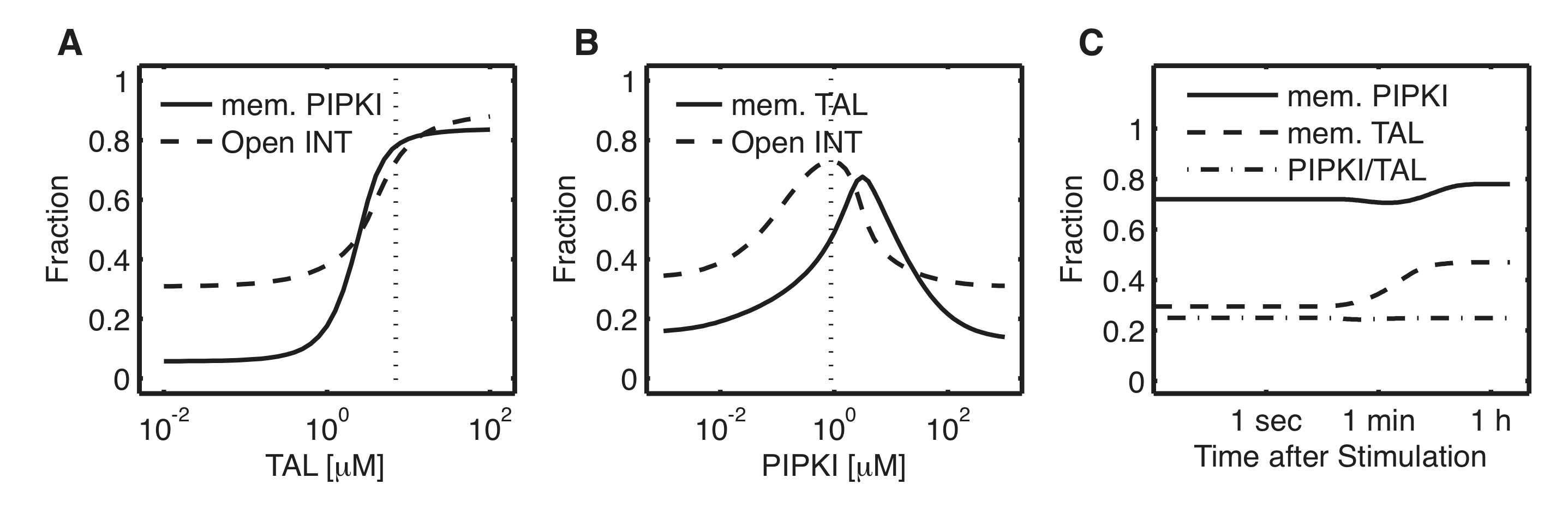}
\end{center}
{\bf Figure 5}

\newpage 

\begin{center}
\includegraphics[width=0.9\textwidth]{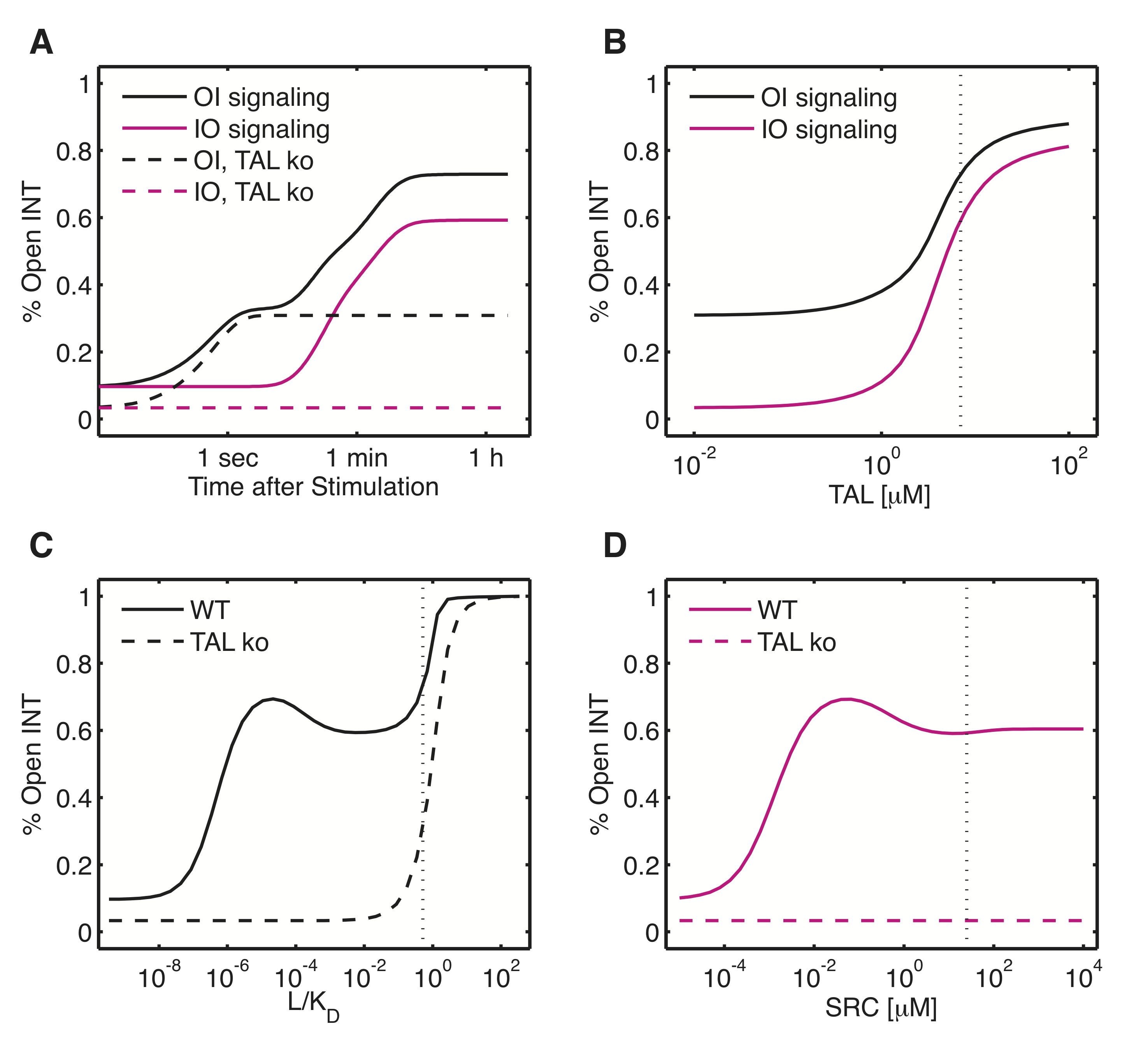}
\end{center}
{\bf Figure 6}
 
\newpage
 
\section*{Tables}
\begin{sidewaystable}[h!]
%\begin{table}[h!]
\small{
\begin{center}
\begin{tabular}{|p{0.025\textwidth}|p{0.2\textwidth}|p{0.25\textwidth}|p{0.08\textwidth}|p{0.45\textwidth}|} \hline
\textbf{Rule} & \textbf{Reaction} & \textbf{Rate [Unit]} & \textbf{Sampled range (x$-$fold)} & \textbf{Reference }\\ \hline
1 & INT opening/closing & $KD1 = 0.03$ \umol, $k1_{-} = 10$ \usec & 2, 10 & \cite{Tadokoro2003, Pampori1999} \\ \hline	
2 & INT/Ligand binding & $KD2 = 0.3$ \umol, $k2_{-} = 0.3$ \usec & fixed, 10 & \cite{Suehiro2000,Faull1993, Iber2006} \\ \hline
3a & INT/TAL binding & $KD3a = 0.1$ \umol, $k3a_{-} = 0.005$ \usec & 2, 10 & \cite{Yan2001,Calderwood2002} \\ \hline	
3b & INT/DOK binding & $KD3b =  10$ \umol, $k3b_{-} = 0.1$ \usec & 2, 10 &\\  \cline{1-4}		
4a & phos.~INT/TAL binding & $2\times KD3a, 2\times k3a_{-}$ & 2 &  \cite{Oxley2008} \\  \cline{1-4}	
4b & phos.~INT/DOK binding & $KD3b/400, k3b_{-}/400$ & 2 & \\ \hline	
5 & SRC opening/closing & $KD5 = 100$ \umol, $k5_{-} = 100$ \usec & fixed, 10 & \cite{Roskoski2004} \\ \hline
6a & SRC/INT binding & $KD6 = 7.5$ \umol, $k6_- = 7.5$ \usec & 5, 10 & \\  \cline{1-4}
6b & SRC/PIPKI binding & \multicolumn{2}{c|}{same as rule 6a} & \\  \cline{1-4}	
7a & integrin dependent SRC phosphorylation & $k7 = 10$ \usec & 5 & \\ \cline{1-4}
7b & cross-talk dependent SRC phosphorylation & \multicolumn{2}{c|}{same as rule 7a; only active in case of outside-in signaling} &  \cite{Songyang1995,Arias-Salgado2003} \\ \cline{1-4}	
8 & PIPKI phosphorylation & $k8 = 10$ \usec & 5 & \\\cline{1-4}	
9 & INT phosphorylation & $k9 = 10$ \usec & 5 &  \\ \hline	
10 & TAL/PIPKI binding & $KD10 = 0.17$ \umol, $k10_- = 0.17$ \usec & 2, 10 & \cite{dePereda2005,Barsukov2003} \\ \hline	
11a & PIPKI dep.~TAL activation & $k11a = 0.1$ \usec & 5 & \\  \hline
11b & PIPKI indep.~TAL activation & $k11b = 0.001$ \usec & 5 & \\ \hline	
12 & TAL deactivation & $k12 = 0.09$ \usec & 2 & \cite{Goksoy2008} \\ \hline	
13 & SRC dephosphorylation & $k13 = 0.001$ \usec & 5 & \\ \cline{1-4}	
14 & PIPKI dephosphorylation & $k14 = 0.001$ \usec & 5 &  \\ \cline{1-4}	
15 & INT dephosphorylation  & $k15 = 0.001$ \usec & 5 & \\ \hline	
16a & TAL membrane shuttling & $KD16a = 0.35$ \umol, $k16a_{-} = 0.35$ \usec & 2, 10 & \cite{Goldmann1995} \\\cline{1-4}	
16b & TAL/PIPKI dimer shuttling & \multicolumn{2}{c|}{same as rule 16a} &  \\ \cline{1-4}	
16c & DOK shuttling & \multicolumn{2}{c|}{same as rule 16a} & \\ \hline	
17a & PIPKI membrane shuttling & $KD17a = 1$ \umol, $k17a_{-} = 1$ \usec & 5, 10 & \cite{Giudici2006} \\ \hline	
17b & TAL/PIPKI trimer shuttling & $KD17b = 0.01$ \umol, $k17b_{-} = 0.01$ \usec & 5, 10 & \\ \hline	
- & $INT_{tot}$ & 40 \umol & fixed & \cite{Neff1982,Akiyama1985,Wiseman2004} \\ \hline
- & $L_{tot}$ & 15 \umol & fixed & \cite{Vitte2004} \\ \hline
- & $TAL_{tot}$ & 1 \umol & 10 & \\ \hline
- & $DOK_{tot}$ & 1 \umol & 10 & \\ \hline
- & $PIPKI_{tot}$ & 0.5 \umol & 10 & \\ \hline
- & $SRC_{tot}$ & 25 \umol & 10 & \\ \hline
\end{tabular}
\end{center}
\caption{{\bf Biochemical reactions rules, parameter values and uncertainty.} Reversible reactions are parameterized by an equilibrium constant and an off-rate; the on-rate is calculated as $KD\cdot k_{-}$. Irreversible reactions have a single reaction rate constant. The uncertainty in the parameter values is indicated by a fold change. }
\label{tab:par}}
\end{sidewaystable}

\clearpage
\newpage

 \setcounter{page}{1}
\section*{Supplementary information}

\subsection*{Rule-based model: molecule types and total concentrations}

\paragraph{Ligand} The concentration of the ligand molecule defines the density of integrin binding sites. However, the decisive quantity in our model is the binding site density relative to the binding site affinity: $L_{tot}/K_D$. We have explored ranges between $10^{-10} \le L_{tot}/K_D \le 10^4$ and used a value of 50 for all time course simulations. Given a $K_D = 0.3 \mu$M, this would correspond to ligand concentration of $L_{tot} = 15 \mu$M, which is among experimentally used ligand densities \citeSupp{Vitte2004, Cox2001}.

\begin{tabular}{p{0.1\textwidth}p{0.75\textwidth}}
\\
{\tt L(I)} & \\
\\
{\tt I} & single integrin binding site  
\end{tabular}

\paragraph{Integrin} There are typically about $5 \times 10^5$ integrins on a fibroblast \citeSupp{Neff1982, Akiyama1985} and about $1-30 \times 10^5$ integrins on CHO cells \citeSupp{Wiseman2004}. The surface area of CHO cells has been measured as 1000-2500 $\mu$m$^2$ \citeSupp{Wiseman2004}. Based on these measurements we expect a density of about 40-3000 integrins per  $\mu$m$^2$ and we shall use 500 integrins per  $\mu$m$^2$ as this corresponds best to both the measurements in fibroblast and CHO cells. If we assume 20 nm as the integrin-ligand binding distance then a density of 500 integrins per  $\mu$m$^2$ corresponds to a concentration of about 40 $\mu$M. 

\begin{tabular}{p{0.1\textwidth}p{0.75\textwidth}}
\\
{\tt INT(Conf\ti O\ti C,L,S1,NPXY\ti U\ti P)} & \\
\\
{\tt Conf\ti O\ti C} & Open and closed integrin conformation \\  
{\tt L} & ligand binding domain \\
{\tt S1} & SRC/FAK beta3 cytoplasmic tail binding domain \\
{\tt NPXY\ti U\ti P} & Talin/Dok binding domain \citeSupp{Calderwood2002}; gets phosphorylated by active Src \citeSupp{Anthis2009}
\end{tabular}

\paragraph{Dok1} \ \\

\begin{tabular}{p{0.1\textwidth}p{0.75\textwidth}}
\\
{\tt DOK1(PTB)}
\\
{\tt PTB} & Integrin binding domain
\end{tabular}

\paragraph{Talin1} \ \\
 
\begin{tabular}{p{0.1\textwidth}p{0.75\textwidth}}
\\
{\tt TAL1(F2,F3\ti O\ti C)} 
\\
{\tt F2} & binding domain for PIP2 (membrane recruitment of talin) \citeSupp{Saltel2009}.  \\
{\tt F3\ti O\ti C} & PTB domain, bind integrin tails and PIPKI, is shielded in closed conformation (autoinhibition) \citeSupp{Ling2003, dePereda2005, Moser2009}. 
\end{tabular}

\paragraph{Src kinase} We have shown previously that the experimentally observed kinetics of Src kinase activation and deactivation can best be reconciled with a Src density of 300 per $\mu m^2$ which corresponds to a Src kinase concentration of about 25 $\mu$M and if Src kinases can interact with all integrin conformations (Felizzi and Iber, submitted).  \ \\

{\tt SRC(SH2,SH3,Y419\ti U\ti P)} \\

\begin{tabular}{p{0.1\textwidth}p{0.75\textwidth}}
{\tt SH2} & interaction domain with CSK \\
{\tt SH3\ti U\ti P} & domain can directly interact with beta3 cytoplasmic tails
\end{tabular}

\paragraph{PIPKI$\gamma$} \ \\

{\tt PIPKI(SPLH,Y644\ti U\ti P)}\\

\begin{tabular}{p{0.1\textwidth}p{0.75\textwidth}}
{\tt Y644\ti U\ti P} & phosphorylted by SRC \citeSupp{Ling2003} \\
{\tt SPLH} & binding domain to F3 domain of talin, Y644 regulated 
\end{tabular}

\subsection*{Rule-based model: reaction rules, parameter values and fold changes}

\paragraph{Rule 1a: Integrin opening and closure} Integrins are allosteric proteins that exist in many conformations  \citeSupp{Hynes2002}. For simplicity we only consider the two extreme conformations, a closed, inactive conformation and an open, active conformation. The conformational equilibrium constant $K_c$ can be derived from the observation that in the absence of ligand only about $2-3 \%$ of all integrins are in the open, high affinity conformation  \citeSupp{Tadokoro2003}. The underlying transition rates have not been measured but a reasonable time scale is $1/10-1$ sec. 
{\tt INT(Conf\ti C,L,NPXY\ti ?) <-> INT(Conf\ti O,L,NPXY\ti ?) k1on, k1off}

\paragraph{Rule 1b: Integrin closure} Applies to all open-DOK-bound-integrins irrespective of the phosphorylation state. \\
{\tt INT(Conf\ti O,L,NPXY\ti ?!1).DOK(BD!1) -> INT(Conf\ti C,L,NPXY\ti ?) + DOK(BD) k1off}

\paragraph{Rule 2: Ligand binding} The dissociation constants for $\alpha_V \beta_3$ integrins in the open conformation has been measured as $K_d \sim 0.3 \mu$M \citeSupp{Plow1981, Faull1995}.  As discussed in detail elsewhere \citeSupp{Iber2006} the affinity of integrins for surface-bound ligand is much lower than for ligand in solution \citeSupp{Kuo1993, Moy1999, Bell1978}. Accordingly we use 500-fold lower affinity constant. For such a lower affinity ligand binding is limited, even at the high ligand densities that are employed in experiments. We used $k_{on} = 1 \mu$M$^{-1}$ s$^{-1}$ as on-rate and we set the off-rate according to the dissociation constant. \\
{\tt INT(Conf\ti O,L) + L(I) <-> INT(Conf\ti O,L!1).L(I!1) k2on, k2off}

\paragraph{Rule 3a: Interaction of unphosphorylated integrin and talin} Talin binds to unphosphorylated integrin $\beta$ tails with at a very low rate, $k3aon = 5 \times 10^4$ M$^{-1}$ s$^{-1}$.  The off-rate was measured as $k3aoff = 5 \times 10^{-3}$ s$^{-1}$ so that $K_d = 0.1 \mu$M \citeSupp{Calderwood2002}. \\
{\tt TAL(Loc\ti PM,Conf\ti O,BD) + INT(Conf\ti O,NPXY\ti U) <-> }\\
{\tt TAL(Loc\ti PM,Conf\ti O,BD!1).INT(Conf\ti O,NPXY\ti U!1) k3aon, k3aoff}

\paragraph{Rule 3b: Interaction of unphosphorylated integrin and Dok} \ \\
{\tt DOK(Loc\ti PM,BD) + INT(Conf\ti O,NPXY\ti U) <-> } \\
{\tt DOK(Loc\ti PM,BD!1).INT(Conf\ti O,NPXY\ti U!1) k3bon, k3boff}

\paragraph{Rule 4a: Interaction of phosphorylated integrin and talin} Phosphorylation of the integrin beta tail reduces the affinity of binding 2-fold \citeSupp{Oxley2008}. \\
{\tt TAL(Loc\ti PM,Conf\ti O,BD) + INT(Conf\ti O,NPXY\ti P) <-> } \\
{\tt TAL(Loc\ti PM,Conf\ti O,BD!1).INT(Conf\ti O,NPXY\ti P!1) k3aon, 2*k3aoff}

\paragraph{Rule 4b: Interaction of phosphorylated integrin and Dok} \citeSupp{Oxley2008} \\
{\tt DOK(Loc\ti PM,BD) + INT(Conf\ti O,NPXY\ti P) <-> } \\
{\tt DOK(Loc\ti PM,BD!1).INT(Conf\ti O,NPXY\ti P!1) k3bon, k3boff/400}

\paragraph{Rule 5: Src opening and closure} In the resting state Src kinases are predominantly inactive ($90-95 \%$) \citeSupp{Roskoski2004}. To obtain such a conformational equilibrium we set our rates for the opening and clamping to $k5on = 1 s^{-1}$ and $k5off = 100 s^{-1}$ respectively. \\
{\tt SRC(Conf\ti C,SH3,Y419\ti U) <-> SRC(Conf\ti O,SH3,Y419\ti U) k5on, k5off}

\paragraph{Rule 6a: Src integrin interaction} The concentration at which half-maximal binding of Src kinases and integrins is achieved has been determined as $EC_{50} \sim 5-10 \mu$M \citeSupp{Arias-Salgado2003} and accordingly we use as dissociation constant $K_d = 7.5 \mu$M for the Src kinase-integrin interaction. \\
{\tt SRC(Conf\ti O,SH3) + INT(BD) <-> SRC(Conf\ti O,SH3!1).INT(BD!1) k6on, k6off}

\paragraph{Rule 6b: Src PIPKI interaction} \ \\
{\tt SRC(Conf\ti O,SH3) + PIPKI(Loc\ti PM,S1) <-> SRC(Conf\ti O,SH3!1).PIPKI(Loc\ti PM,S1!1) k6on, k6off}

\paragraph{Rule 7a: integrin-dependent Src phosphorylation} Activation of Src kinases is achieved by trans autophosphorylation on Tyr-418 \citeSupp{Harrison2003} and thus requires juxtaposition of open Src kinases, presumably by binding to ligand-bound integrins \citeSupp{Arias-Salgado2003}. We do not model juxtaposition of integrins explicitly, but require ligand bound integrins. \\
{\tt SRC(SH3!1,Y419\ti U).INT(BD!1,L!+) -> SRC(SH3!1,Y419\ti P).INT(BD!1,L!+) k7}

\paragraph{Rule 7b: cross-talk dependent Src phosphorylation} Inside-out signaling is simulated by setting IOFLAG=1 and $L_{tot} = 0$.\\
{\tt SRC(SH3!1,SH3,Y419\ti U) -> SRC(SH3!1,SH3,Y419\ti P) IOFLAG*k7}

\paragraph{Rule 8: PIPKI phosphorylation} The $K_m$ for the optimal Src kinase substrate has been determined as $K_m = 30\mu$M, and accordingly we use $k_{cat} = 10$ s$^{-1}$ \citeSupp{Songyang1995}. In the absence of experiments that would suggest otherwise, we choose the same $K_m$ and $k_{cat}$ values for all Src-mediated reactions, i.e. also for the phosphorylation of integrin $\beta$ tails and of PIPKI. \\
{\tt PIPKI(S1!1,Y644\ti U).SRC(SH3!1,Y419\ti P) -> PIPKI(S1!1,Y644\ti P).SRC(SH3!1,Y419\ti P) k8}

\paragraph{Rule 9: Integrin phosphorylation} Integrins have to be in an open conformation, SRC bound, but not necessarily bound by ligand. \\
{\tt SRC(SH3!1,Y419\ti P).INT(BD!1,NPXY\ti U) -> SRC(SH3!1,Y419\ti P).INT(BD!1,NPXY\ti P) k9}

\paragraph{Rule 10: Tal and PIPKI interaction} In the quiescent state most talin cannot bind integrins because of self-inhibitory interactions \citeSupp{Goksoy2008}. The inhibited forms can, however, still bind PIPKI. The affinity between PIPKI and talin has been determined as 170 nM  \citeSupp{dePereda2005,Barsukov2003}. Since PIPKI exists as a dimer, we model two talin binding sites leading to TAL:PIPKI dimers and trimers. Without loss of generality we assume a sequential binding. \\
{\tt TAL(Loc\ti PM,BD) + PIPKI(Loc\ti PM,T1,T2) <-> }\\
{\tt TAL(Loc\ti PM,BD!1).PIPKI(Loc\ti PM,T1!1,T2) k10on, k10off} \\
{\tt TAL(Loc\ti PM,BD) + PIPKI(Loc\ti PM,T1!+,T2) <-> } \\
{\tt TAL(Loc\ti PM,BD!1).PIPKI(Loc\ti PM,T1!+,T2!1) k10on, k10off} \\
{\tt TAL(Loc\ti CP,BD) + PIPKI(Loc\ti CP,T1,T2) <-> } \\
{\tt TAL(Loc\ti CP,BD!1).PIPKI(Loc\ti CP,T1!1,T2) k10on, k10off} \\
{\tt TAL(Loc\ti CP,BD) + PIPKI(Loc\ti CP,T1!+,T2) <-> } \\
{\tt TAL(Loc\ti CP,BD!1).PIPKI(Loc\ti CP,T1!+,T2!1) k10on, k10off}

\paragraph{Rule 11a: Activation of talin} Most cellular talin is unable to bind integrins because of inhibitory self-interactions \citeSupp{Goksoy2008}. Talin can be activated by PIPKI which produces PIP2 \citeSupp{DiPaolo2002,Ling2002}. PIP2 is highly unstable \citeSupp{Hilgemann2007}, and its production, decay and diffusion are not modelled explicitly. We rather assume that talin must be bound to membrane-bound, active PIPKI for activation.  The rate of PIPKI-dependent PIP2 production and talin activation has not been measured, and we use $k11a = 0.1 s^{-1}$. \\
{\tt TAL(Loc\ti PM,Conf\ti C,BD!1).PIPKI(Loc\ti PM,T1!1,Y644\ti P) -> } \\
{\tt TAL(Loc\ti PM,Conf\ti O,BD!1).PIPKI(Loc\ti PM,T1!1,Y644\ti P) k11a } \\
{\tt TAL(Loc\ti PM,Conf\ti C,BD!1).PIPKI(Loc\ti PM,T2!1,Y644\ti P) -> } \\
{\tt TAL(Loc\ti PM,Conf\ti O,BD!1).PIPKI(Loc\ti PM,T2!1,Y644\ti P) k11a}

\paragraph{Rule 11b: Activation of talin} We also assume a PIPKI independent activation of talin at the membrane with a very low rate of $k11b = 10^{-3}$.\\
{\tt TAL(Loc\ti PM,Conf\ti C) -> TAL(Loc\ti PM,Conf\ti O) k11b}

\paragraph{Rule 12: Inactivation of talin} Only if it is unbound \\
{\tt TAL(Conf\ti O,BD) -> TAL(Conf\ti C,BD) k12}

\paragraph{Rule 13: Src de-phosphorylation} We include constitutive Src, integrin, and PIPKI dephosphorylation at rates 0.1 s$^{-1}$. These rates have not been measured directly in experiments, but data exist that provide some bounds. Thus in talin knock-out cells Src activity drops to 50 $\%$ of its maximal value within 15-20 minutes. For large dephosphorylation rates there is little Src activation and the system rapidly equilibrates. For small dephosphorylation rates rapid Src deactivation is impossible. The rate of Src dephosphorylation must therefore be about $k13 = 0.1 s^{-1}$. \ \\
{\tt SRC(Conf\ti O,Y419\ti P) -> SRC(Conf\ti O,Y419\ti U) k13}

\paragraph{Rule 14: PIPKI de-phosphorylation} Must be unbound by Src \\
{\tt PIPKI(Y644\ti P,S1) -> PIPKI(Y644\ti U,S1) k14}

\paragraph{Rule 15: integrin dephosphorylation} Must be unbound by Src \\
{\tt INT(NPXY\ti P,BD) -> INT(NPXY\ti U,BD) k15}

\paragraph{Rule 16a: talin shuttling} \ \\
{\tt TAL(Loc\ti CP,BD) <-> TAL(Loc\ti PM,BD) k16on, k16off}

\paragraph{Rule 16b: PIPKI shuttling} incase of dimers \\
{\tt TAL(Loc\ti CP,BD!1).PIPKI(Loc\ti CP,T1!1,T2,S1) <-> } \\
{\tt TAL(Loc\ti PM,BD!1).PIPKI(Loc\ti PM,T1!1,T2,S1) k16on, k16off }

\paragraph{Rule 16c: DOK shuttling} (same parameters as for talin) \\
{\tt DOK(Loc\ti CP,BD) <-> DOK(Loc\ti PM,BD) k16on, k16off}

\paragraph{Rule 17a: PIPKI shuttling} Phosphatidylinositol phosphate kinase type I$\gamma$ (PIPKI) appear not to bind well to membrane lipids and PIPKI is therefore assumed to be homogeneously distributed between membrane compartment and cytoplasm \citeSupp{Giudici2006}, i.e PIPKI shuttles between membrane and cytoplasm at rate 1 s$^{-1}$ in both directions. \\
{\tt PIPKI(Loc\ti CP,T1,T2,S1) <-> PIPKI(Loc\ti PM,T1,T2,S1) k16on, k17offA}

\paragraph{Rule 17b: PIPKI shuttling} in case of trimers \\
{\tt TAL(Loc\ti CP,BD!1).PIPKI(Loc\ti CP,T1!1,T2!2,S1).TAL(Loc\ti CP,BD!2) <-> } \\
{\tt TAL(Loc\ti PM,BD!1).PIPKI(Loc\ti PM,T1!1,T2!2,S1).TAL(Loc\ti PM,BD!2) k16on, k17offB}

\section*{Compartment modeling}

We consider two reaction compartments: cytoplasm (CP) and plasma membrane (PM). TAL, PIPKI and DOK can shuttle between both compartments. Since the CP volume is about 20 fold larger compared with the effective PM reaction volume, the shuttling rates have to account for the change in species concentration due to shuttling. Therefore, we normalize the compartment exchange fluxes on a per-species basis as detailed below. Note, that this approach differs from the standard SBML formulation, which relies on an extensive description of species levels (i.e. species levels are treated as absolute numbers and not as concentrations) \citeSupp{Hucka2003}.

Let $A$ and $B$ be two compartments with reaction volume $V_A$ and $V_B$, respectively. Let $n_A$ be the number of particles in compartment $A$ and $n_B$ the number of particles in compartment $B$. For simplicity we assume a constant number of particles.
\begin{eqnarray*}
n_T & = n_A + n_B
\end{eqnarray*}
The concentrations in the compartments are denoted as $x_A = n_A/V_A$  and $x_B = n_B/V_B$. We want to describe the change in concentration due to the flux of particles between the compartments. We assume large particle numbers, such that the problem can be formulated in terms of a set of differential equations.
\begin{eqnarray}
\dot x_A & = - \alpha_{A\rightarrow B} + \alpha_{B\rightarrow A} \\
\dot x_B & = - \beta_{B\rightarrow A} + \beta_{A\rightarrow B} 
\end{eqnarray}
We will assume mass-action kinetics, i.e., all fluxes are proportional to the concentrations of the reacting species. 
\begin{eqnarray*}
\alpha_{A\rightarrow B} = a_1 x_A\\
\alpha_{B\rightarrow A} = a_2 x_B\\
\beta_{B\rightarrow A} = b_1 x_B\\
\beta_{A\rightarrow B} = b_2 x_A
\end{eqnarray*}
Due to conservation of particles $\dot n_A + \dot n_B = 0$, and therefore
\begin{eqnarray*}
\dot x_A V_A  + \dot x_B V_B  =  -a_1 n_A + a_2 \frac{V_A}{V_B} n_B  - b_1 n_B + b_2 \frac{V_B}{V_A} n_A \stackrel{!}{=}  0.
\end{eqnarray*}
The last equation can be rearranged to 
\begin{eqnarray*}
n\cdot v = 0, 
\end{eqnarray*}
where $n = (n_A,n_B)$ and $v = (b_2 \frac{V_B}{V_A}  - a_1, a_2 \frac{V_A}{V_B} -  b_1)^T$. This equation must hold for all $n$. As an example, consider the two equations $n_1 \cdot v = 0$ and $n_2 \cdot v = 0$ for linearly independent vectors $n_1$ and $n_2$, which can only hold true for $v=(0,0)$. This is therefore the general solution for arbitrary $n$. It follows
\begin{eqnarray*}
b_1 & = a_2 \frac{V_A}{V_B} \\
b_2 & = a_1 \frac{V_A}{V_B} .
\end{eqnarray*}
Inserting these into the original equation for $x_B$ leads to  
\begin{eqnarray*}
\dot x_B & = - (a_1 V_A x_B + a_2 V_A x_A ) / V_B.
\end{eqnarray*}
Defining the new parameters $\tilde{a_1} = a_1 V_A$ and $\tilde{a_2} = a_2 V_A$, we can rewrite the original equations in terms of the new parameters as
\begin{eqnarray}
\dot x_A & = (- \tilde{a_1} x_A + \tilde{a_2} x_B)/V_A \\
\dot x_B & = (- \tilde{a_2} x_B + \tilde{a_1} x_A)/V_B .
\end{eqnarray}
The dimension of $\tilde{a_1}$ and $\tilde{a_2}$ are volume over time and are explicitly referencing $V_A$. However, we can by the same approach define parameters  $\tilde{b_1} = b_1 V_B$ and $\tilde{b_2} = b_2 V_B$ referencing volume $V_B$ and derive the equations in terms of $\tilde{b_1}$ and $\tilde{b_2}$.
\begin{eqnarray*}
\dot x_A & = (- \tilde{b_2} x_A + \tilde{b_1} x_B)/V_A \\
\dot x_B & = (- \tilde{b_1} x_B + \tilde{b_2} x_A)/V_B 
\end{eqnarray*}
From this it is apparent that $\tilde{a_1} = \tilde{b_2} \equiv k_{A\rightarrow B}$ and $\tilde{a_2} = \tilde{b_1} \equiv k_{B\rightarrow A}$.

Particle numbers might not be conserved due to production or decay occurring within compartments. However, in mass-action kinetics velocities are proportional to concentrations: changing the reaction volume while keeping the concentrations fixed, does not change the velocity of a reaction. Therefore, the transformations $a_1 \rightarrow \tilde{a_1}$ and $a_2 \rightarrow \tilde{a_2}$ do not affect other fluxes within the compartments, whose rates are volume independent. Therefore the general form of the reaction kinetics including production, decay and compartment exchange is: 
\begin{eqnarray}
\dot x_A & = (- \tilde{a_1} x_A + \tilde{a_2} x_B)/V_A + f_{prod}(x_A) - f_{deg}(x_A)\\
\dot x_B & = (- \tilde{a_2} x_B + \tilde{a_1} x_A)/V_B + f_{prod}(x_B) - f_{deg}(x_B).
\end{eqnarray}

\section*{Evaluation of sample size}

Exhaustive sampling of the biologically feasible parameter range is computationally infeasible. Since most of the parameter combinations lead to qualitatively similar model dynamics we are interested in the effective number of samples needed to get a comprehensive picture of the possible model dynamics. To this end we monitor the convergence of the time-dependent mean of the model observables for an increasing number of samples by a blocking procedure. Samples of size $N = \{10,10^2,10^3,10^4,10^5\}$ are divided into 10 blocks. For each sample size we calculate the overall mean as well as the standard deviation of the means of all 10 blocks. Next, we compute the coefficient of variation (CV), i.e., the standard deviation divided by the mean, for each sample size. This gives us a CV value for each sample size, time point and model observable (\fig{fig:ssize}). The analysis shows, that the average signaling dynamics is sufficiently well captured with $10^5$ samples, i.e. $CV<0.1$ for all observables and time points.

\newpage
\bibliographystyleSupp{/Users/iberd/MyDocuments/Academia/Publications/Bibliography/msb}
\bibliographySupp{integrin.bib}

\newpage
\renewcommand{\thefigure}{S\arabic{figure}}
\setcounter{figure}{0}
\begin{figure}[h!]
\begin{center}
\includegraphics[width=1\textwidth]{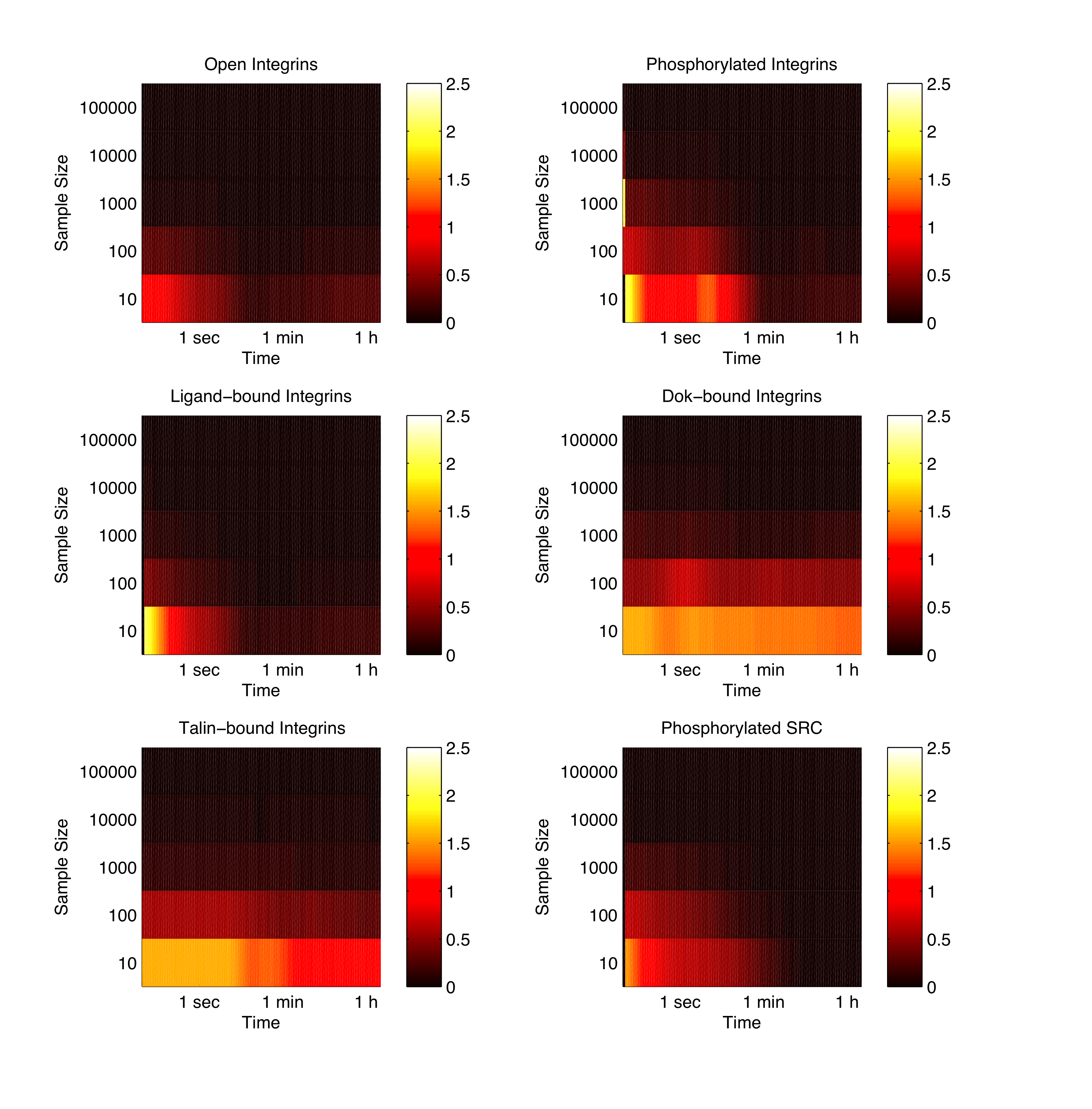}
\end{center}
\caption{{\bf Mean signaling dynamics in dependence of the sample size.} Each plot shows the coefficient of variation (standard deviation/mean) in dependance of time (horizontal axis) and sample size ( vertical axis) for a selected observable. Mean and standard deviation are calculated by a blocking procedure as described in the text. For a sample size of $10^5$ the coefficient of variation is below 0.1 \% for all time points and observations.}
\label{fig:ssize}
\end{figure}

\newpage

\renewcommand{\thetable}{S\arabic{table}}
\setcounter{table}{0}
\begin{table}[t]
\begin{center}
\begin{tabular}{l|r|r}
Parameter & Group 1 & Group 2 \\ \hline
k1on & 0.68 & 0.4 \\ \hline 
k1off & 19.749 & 10.467 \\ \hline 
k2on & 2.353 & 0.950 \\ \hline 
k2off & 0.706 & 0.285 \\ \hline 
k3aon & 0.106 & 0.117 \\ \hline 
k3aoff & 0.009 & 0.014 \\ \hline 
k3bon & 0.018 & 0.055 \\ \hline 
k3boff & 0.191 & 0.609 \\ \hline 
k5on & 2.154 & 2.128 \\ \hline 
k5off & 313.8 & 319.7 \\ \hline 
k6on & 2.107 & 2.093 \\ \hline 
k6off & 23.87 & 24.26 \\ \hline 
k7 & 14.89 & 15.09 \\ \hline 
k8 & 14.85 & 15.21 \\ \hline 
k9 & 14.89 & 15.07 \\ \hline 
k10on & 3.423 & 2.042 \\ \hline 
k10off & 0.666 & 0.374 \\ \hline 
k11a & 0.214 & 0.143 \\ \hline 
k11b & 0.0006 & 0.0015 \\ \hline 
k12 & 0.093 & 0.1 \\ \hline 
k13 & 0.002 & 0.002 \\ \hline 
k14 & 0.002 & 0.001 \\ \hline 
k15 & 0.0015 & 0.0014 \\ \hline 
k16on & 2.071 & 2.075 \\ \hline 
k16off & 0.704 & 0.692 \\ \hline 
k17offA & 2.839 & 2.902 \\ \hline 
k17offB & 0.03 & 0.033 \\ \hline 
$INT_{tot}$ & 40 & 40 \\ \hline 
$L_{tot}$ & 15 & 15 \\ \hline 
$TAL_{tot}$ & 6.98 & 0.68 \\ \hline 
$DOK_{tot}$ & 1.74 & 6.05 \\ \hline 
$PIPKI_{tot}$ & 0.88 & 1.38 \\ \hline 
$SRC_{tot}$ & 25 & 25 
\end{tabular}
\end{center}
\caption{Mean parameter values of group 1 (high TAL:INT, Figure 3 main text) and group 2 (high DOK:INT, Figure 3 main text). On-rates are related to the KD value mentioned in the main text as $kon = koff/KD$.}
\label{tab:parS}
\end{table}

\end{document}